# AI Tools Adoption across the Double Diamond Workflow: Phase, Mode, and Barriers in Designer Practice

**Sepideh Tajarmakan** *

[1]Independent Researcher

m.s.tajarmakan@gmail.com
ORCID 0009-0009-3418-8893

**Khashayar Hojjati Emami**[2]

[2]Iran University of Art, Department of Industrial Design

k.emami@art.ac.ir
ORCID 0000-0002-7325-8091

** Corresponding author*

**ABSTRACT**

Designers are adopting AI faster than the tools built for them can keep up. General-purpose conversational AI led designers' AI use in three of the four phases of design work, ahead of tools built for design. We surveyed 443 designers from at least 43 countries about AI use in each phase of the Double Diamond: Discover, Define, Develop and Deliver. Tools were classified by output and function. Of these designers, 79.7% confirmed AI use in at least one phase. Adopters used AI in a mean of 2.89 phases, and only 35.1% used it in all four. They added phases in workflow order, from Discover onward. Conversational tools led in Discover, Define and Deliver. AI-native image generation led in Develop, where AI-native tools peaked at 57.6% of tool picks. Embedded AI features in existing design software peaked later, in Deliver (37.5%). Fewer designers used AI in later phases, from 318 in Discover to 182 in Deliver. Those who did named a similar number of tools per phase. Low perceived usefulness was the only barrier that separated non-adopters (40.5%) from adopters (16.4%) after correction for multiple tests. Concern about accuracy was similar in both groups.

# INTRODUCTION

Designers now use AI tools for ideation, content generation, visual exploration and workflow support (Aphirakmethawong et al., 2022; Durgam et al., 2025). Design is complex work made of interdependent activities (Retelny et al., 2017). The work also depends on the designer's role, the task context and the phase of the design process. Knowing where designers adopt AI can show which tools fit which stage (Suchman, 1987; Orlikowski, 2000; Dourish, 2001). This study uses the Design Council's Double Diamond to structure the analysis of AI use across the design process. Its phases are Discover, Define, Develop and Deliver (Design Council, 2007). Discover centers on research and open exploration, such as brainstorming and user research. Define turns what Discover found into a clear problem statement. Develop generates and prototypes possible solutions. Deliver tests, refines and finalizes the chosen solution. The Double Diamond is a widely adopted model for mapping AI onto the design process (Zhou & Chen, 2025; Mirzaei et al., 2026). This model also fits UX/UI, product, industrial and graphic design. Existing research leaves five gaps that this study addresses.

Empirical work is largely small-sample, with six to 21 interviews or single workshops, or literature-derived (Saadi & Yang, 2023; Hamilton et al., 2024; Mirzaei et al., 2026). Two recent large sample studies begin to address this. One measures acceptance intention rather than reported behavior (Zou et al., 2026). The other compares adoption across national contexts using a structured acceptance model (Fang et al., 2026). Neither study reports adoption as behavior disaggregated by phase and tool, which is the gap our study addresses. "AI use" is typically conceptualized as a general behavior, without specific tool, output type or function (Y. Lu et al., 2024). Even where tool-level detail exists, coverage is uneven across the workflow (Veggi, 2025). Frameworks mapping AI onto design phases exist largely at a conceptual level (Lee et al., 2025; Wehnert & De Luca, 2024). They are not grounded in large scale reported behavior by phase. Little is known about how much AI is used beyond presence or absence. Where measured, frequency covers a single tool category, not a full workflow (Tang et al., 2024). Existing work treats non-adoption as a general attitudinal profile (Fang et al., 2026). It does not show what specifically differentiates non-adopters from adopters on the same measure.

We surveyed 443 designers from at least 43 countries and regions. They included UX/UI, Product, Industrial and Graphic designers, students and related roles. Closed and open-ended questions followed each Double Diamond phase and addressed four research questions.

- **RQ1 — Adoption:** What proportion of designers report using AI tools, and how many phases of the workflow does that use typically span?
- **RQ2 — Tool category:** Which categories of AI tools, by output and function, are used within each phase?
- **RQ3 — Intensity:** How does the depth of AI use change across the workflow?
- **RQ4 — Barriers:** What barriers to AI adoption do designers report, and how do these differ between adopters and non-adopters?

"AI tool" here covers two kinds of tools. AI-native applications are standalone generative tools, such as Midjourney, and Embedded AI features sit inside existing design software, such as Photoshop's Generative Fill. The full classification is given in the Method. This study reports AI use by phase, tool type and barrier in one large sample.

## LITERATURE REVIEW

### *AI Tool Definition*

The OECD (2024) defines AI as a machine-based system that infers from its input how to generate predictions, content, recommendations or decisions. These systems vary in autonomy and adaptiveness after deployment. The term is younger than the idea. McCarthy coined "artificial intelligence" at a 1956 workshop (Grudin, 2009). Its boundaries have shifted since, from Turing's test of conversational indistinguishability (Crevier, 1993) to today's generative systems. Definitions still vary, and some set a capability ceiling of narrow, general or super intelligence (Kaplan & Haenlein, 2019). Others frame AI as adaptation under insufficient knowledge and resources (Wang, 2019). AI and HCI developed as separate fields on shared ground. Human-centered AI aims to close this divide by centering people (Grudin, 2009; Auernhammer, 2020; Schmidt, 2020). For HCI, the layer through which people meet AI matters more than its definition. This layer is the AI tool, the application that puts a model to work on a task. This focus follows an older tension in AI framing. McCarthy saw AI as autonomous intelligence and Engelbart as augmentation of human capability (Spallazzo & Sciannamè, 2022; Winograd, 2006). Tools have been classified by internal world-representation (Chassang et al., 2021) and by autonomy relative to human direction (Parasuraman et al., 2000; Shneiderman, 2020). Output type is a third basis, from text and image to video, audio and code (Garrido-Merchán & López, 2023). We classify tools by two observable dimensions, output and function within output. Both describe what a designer produces and are easier to report than internal properties.

### *AI Concentrates Unevenly Across the Design Process*

Structured process frameworks let researchers ask where AI is used, as well as whether it is used. The Double Diamond serves that purpose here. Its four phases map onto the questionnaire's own structure (Design Council, 2007). Some work finds AI concentrated in the latter, more generative stages. In built systems, Develop and Deliver outnumber Discover and Define (Lee et al., 2025). A broader synthesis by Mirzaei et al. (2026) shows AI's role differentiated by stage, with no concentration at either end. Several platforms are now built on the Double Diamond structure, from multi-agent visual design systems (Wen, 2025) to interdisciplinary co-design tools (Tseng & Chang, 2025). Yet this body of work describes what tools are built to do. Our results set it against what designers report doing.

### *Tool Choice and Usage Depth Remain Underspecified*

Much existing literature treats "AI use" as a general behavior and does not examine which specific tools designers reach for (Y. Lu et al., 2024). Where tool-level detail exists, coverage across the workflow is uneven. Veggi (2025) documents strong tool availability for text and image tasks but weak coverage for 3D modeling and evaluation. Named tool patterns for other output types remain largely undocumented. A related gap concerns usage intensity. Tang et al. (2024) found that professional designers used AI image generation tools less often than non-professionals. That study covered one tool category and gave no account of intensity across a full workflow.

### Non-Adoption Is Described as a General Attitude

Research on designers who do not adopt AI has described non-adoption as a general attitude. Few studies have isolated the concerns that separate non-adopters from adopters. One recent profiling study identifies five adopter personas, from “Hesitant Beginners” to “Resistant Experts” (Fang et al., 2026). Each persona carries its own barriers and motivations. The typology is useful, yet it rests on general orientation and does not compare non-adopters with adopters item by item. This study compares barrier endorsement between adopters and non-adopters on the same measure.

## METHOD

### Study Design

This study used a cross-sectional, quantitative survey design to examine how designers report using AI tools across the four phases of the Double Diamond. Data were collected through a confidential, self-administered online questionnaire in Google Forms between August 2025 and February 2026. This design gives a broad, descriptive snapshot of adoption in a large sample spanning at least 43 countries and regions. It favors breadth over tracking change in individual designers across time. The questionnaire also included open-ended items for naming tools not on the preset list, described under Data Cleaning.

### Sample Size

Sample size was estimated a priori with Cochran’s formula for an unknown population proportion (Cochran, 1977). The inputs were $Z = 1.96$ for 95% confidence, $P = 0.50$ for maximum variability and $e = 0.05$ as the margin of error. The formula gave $n_0 \approx 385$. The final sample of 443 valid responses exceeded this threshold.

### Sampling Procedure

We used non-probability sampling that combined purposive, convenience and snowball methods. Recruitment targeted design professionals and students in product, UX/UI, industrial and graphic design roles. An “Other” category covered unlisted specialties such as jewelry, fashion, animation and game design. Experience level and country were recorded but not used for eligibility. The survey was shared through direct LinkedIn outreach, design-focused Facebook and Telegram groups and the authors’ professional networks. Recipients were encouraged to forward it, which supported snowball distribution.

### Questionnaire Structure

Instructions explained the study’s purpose and each section. The questionnaire had 70 items, with consent followed by eight sections (Table 1). Demographic questions came next (Q2 to Q9), with the gender item added after collection began. A screening question on AI tool use followed (Q10). The 84 who answered “No” went to the Ethics and Barriers section. Another 359 answered “Yes”, reported frequency of use (Q11) and step use (Q12), and continued to the phase sections. Each phase opened with a Yes or No question on AI use. Users selected tools from a preset list with an open-text option, and non-users skipped ahead. Six participants who answered "Yes" to the general question did not confirm use in any single phase. Their handling is described under Data Analysis Plan. This study draws on the consent and demographic items (Q1 to Q9) and the general and phase AI use items (Q10 to Q13, Q23, Q33 and Q43). It also draws on the phase tool selection items (Q14, Q24, Q34 and Q44) and the barriers item (Q62).

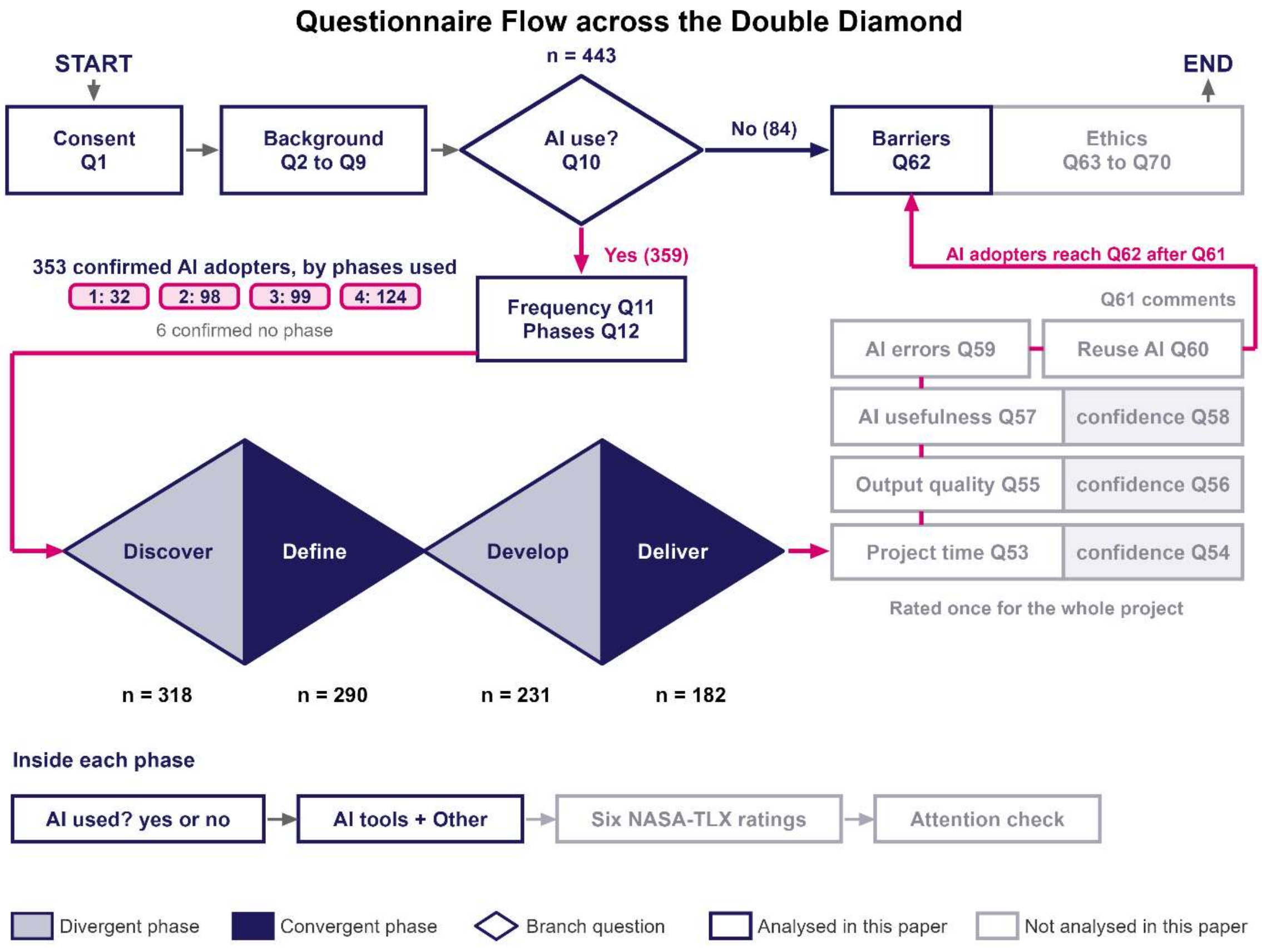

**Figure 1: Questionnaire Flow across the Double Diamond**

**Table 1: Questionnaire Sections and Item Counts**

| Section | Question range | Items |
|---|---|---|
| Consent | Q1 | 1 |
| Demographics | Q2 – Q9 | 8 |
| AI overall | Q10 – Q12 | 3 |
| Discover phase (AI-tools, NASA-TLX) | Q13 – Q22 | 10 |
| Define phase (AI-tools, NASA-TLX) | Q23 – Q32 | 10 |
| Develop phase (AI-tools, NASA-TLX) | Q33 – Q42 | 10 |
| Deliver phase (AI-tools, NASA-TLX) | Q43 – Q52 | 10 |
| KPI | Q53 – Q61 | 9 |
| Ethics and Barriers | Q62 – Q70 | 9 |
| **Total** | | **70** |

***Note: Phase rows are indented under the phase-by-phase section. Each phase block was shown only to participants who answered "Yes" to the general screening question (Q10).***

Consent, demographic, screening and barriers items were answered by all 443 respondents. Phase items were answered only by those who reported AI use in that phase. The phase samples were 318 in Discover, 290 in Define, 231 in Develop and 182 in Deliver. This decline is itself a result, since it follows real adoption behavior. The unequal phase samples still make statistical comparison across phases harder, as discussed under Limitations. Other items covered NASA-TLX workload, KPI outcomes, ethics and risk perception. They serve planned companion analyses and are not analyzed here (Figure 1).

## Data Cleaning and Filtering

The dataset was cleaned by standardizing open-text responses into consistent categories. Tool names were checked for spelling variants, capitalization and duplicate labels. Entries that clearly named the same tool were merged under one canonical name. Distinct features of the same platform, such as Figma AI and Figma Make, were kept separate because they serve different user tasks. Tool names also varied across phase datasets, as with "Bolt.new" and "Bolt" or "Claude" and "Clude". Some tools were rebranded during the study, as when Bing AI became Microsoft Copilot. Before matching against the taxonomy, all names were normalized by removing punctuation, slashes and extra spaces and by standardizing case. Entries not matched to a specific tool were excluded from tool counts. These included generic references, for example "internal company tools", and tools without any AI feature. One entry, "Mira", had several possible referents and was excluded from tool counts. Country responses were cleaned in the same way. Spelling and capitalization were corrected, and city-only responses were recoded to their country where possible. Multi-country responses were combined into one category. After cleaning, 432 participants reported 43 countries and regions, and 11 (2.5%) chose "prefer not to say". No participant was excluded for age, education, experience or role. Country responses and open-text AI tool names were manually reviewed and standardized in Excel. The four phase-specific datasets were then merged, together with the original demographic export, into a single master file using Python (version 3.13; Python Software Foundation (2024); package: pandas) which combined demographic, screening, and tool-selection data without altering any cleaned values. In each phase, open-text "Other" responses were split into single tool mentions, since participants often listed several tools in one entry. Each identified tool became a separate binary column coded 0 or 1, in the same structure as the preset checkboxes. Statistical analyses and visualizations were conducted in R (version 4.5.3; R Core Team (2026); packages: openxlsx, dplyr, tidyr, ggplot2, patchwork, gridExtra, irr). Group comparisons used chi-square tests with Yates correction and McNemar tests with continuity correction. Phase comparisons used the Friedman test with Kendall's W and Wilcoxon signed-rank tests with Holm correction. Spearman correlation and Shapiro-Wilk tests were also run. Open-text data were used only to identify tools. No thematic, narrative or interpretive analysis was conducted, and the study is quantitative throughout.

## AI Tool Classification

Across the study, 73 distinct AI tools were identified: 18 (24.7%) offered directly as preset checkbox options (marked * in Appendix Table 3), plus 55 (75.3%) named by participants through open-text, all coded as binary (0/1) variables per Data Cleaning. The preset checkbox options included: Discover (ChatGPT, Perplexity AI, Elicit, Figma AI, Miro AI, Autodesk Generative

Design, , Adobe Sensei); Define (ChatGPT, Notion AI, FigJam AI, Miro AI, Autodesk Fusion 360, Whimsical AI); Develop (Midjourney, DALL·E, Figma AI, Adobe Firefly, Runway ML, Autodesk Fusion 360); and Deliver (Adobe Photoshop Generative Fill, Canva Magic Studio, Runway ML, ChatGPT, Descript). Selected based on a July 2025 prestudy, these tools served as phase-specific examples, with an Other option allowing participants to report additional tools, as the study focuses on AI use broadly rather than specific platforms. Of these 73 tools, 14 (19.2%) appeared in all four phases, 15 (20.5%) in two, 11 (15.1%) in three, and 33 (45.2%) in only one; indicating a small core of general-purpose tools used throughout, but a tool ecosystem that is mostly phase-specific.

Tools were classified along two dimensions, primary output and function within that output. Output had five categories (Gozalo-Brizuela & Garrido-Merchán, 2023). These were Text, Image and Video, plus 3D Geometry and Agentic. Function captured the tool's role within its output type. For most outputs, it separated AI-native generation from AI features embedded in existing software (Q. Lu et al., 2024). Within Text, it separated conversational assistance, code generation and research synthesis. The resulting 10 function-nests were spread unevenly across the five outputs, as in recent multi-dimensional taxonomies of generative AI systems (Shi et al., 2024; Doshi & Moore, 2026). Agentic tools Manus AI and Marvin, classified per Doshi and Moore's (2026) agentic category, were reported only twice across all four phases combined. They are retained in the full taxonomy for completeness but excluded from phase level interpretation due to insufficient frequency. Some tools support several outputs or functions, such as Krea for image and video generation. These tools were classified by their primary, most established use at data collection, unless a participant's open-text description said otherwise. This rule was applied to every tool, even where no ambiguity was suspected. Of the 73 tools, only 9 (12.3%) fall under Conversational & General-Purpose and can serve many tasks. The other 64 (87.7%) are specialized to a single task or output type.

## *Data Analysis Plan*

Tool counts were right-skewed and zero-inflated. Shapiro-Wilk tests on the full sample ($N = 443$) rejected normality in all four phases (Discover $W = 0.866$, Define $W = 0.803$, Develop $W = 0.748$, Deliver $W = 0.701$, all $p < 0.001$) (Shapiro & Wilk, 1965). Non-parametric methods were therefore used throughout. Group differences used chi-square tests of independence (Pearson, 1900), and phase comparisons used the Friedman test (Friedman, 1937). Post-hoc comparisons used Wilcoxon signed-rank tests (Wilcoxon, 1945) with Holm correction (Holm, 1979), and effect size was reported as r (Cohen, 2013). Each research question draws on a different analytic sample. **RQ1** reports adoption among all 443 participants as confirmed AI use in at least one phase ($n = 353$). It also reports mean breadth of use among adopters, and both measures are descriptive. General frequency of AI use (Q11, $n = 359$) was summarized descriptively. Its association with breadth among adopters ($n = 353$) was tested with Spearman's rank correlation (Spearman, 1904), which suits ordinal, non-normal variables. **RQ2** reports tool use by output modality and function within each phase (Discover $n = 318$, Define $n = 290$, Develop $n = 231$, Deliver $n = 182$). **RQ3** compared tool-count intensity across the four phases at two levels. At the population level ($N = 443$), non-adopters in a phase were coded as zero and phases were compared with a Friedman test. The test is rank-based and does not assume normality, which the Shapiro-Wilk results ruled out.

Effect size was reported as Kendall's W (Kendall & Smith, 1939). The branching design skipped non-adopters in each phase instead of measuring them. A separate descriptive account for adopters only (M, SD, IQR) is therefore reported alongside the population test. After a significant Friedman result, pairs of phases were compared with the Wilcoxon and Holm procedure above. **RQ4** compared barrier endorsement between adopters and non-adopters with chi-square tests of independence. Each of the ten barrier categories was tested separately, with Holm correction across the ten tests and $\varphi$ as effect size. These tests use the general screening question (359 adopters, 84 non-adopters), because the barriers item followed that question in the branching logic. The phase-confirmed count used in RQ1 ($n = 353$) was not used here.

## Validity and Reliability

Validity and reliability are reported with the four-part framework of Runeson and Höst (2009). Its parts are construct validity, internal validity, external validity and reliability. Internal validity does not apply, because this study is descriptive and not causal. Reported associations, such as those between phase and AI-use intensity, should not be read as causal mechanisms. **External validity** is limited by non-probability snowball sampling and the resulting concentration of the sample in Iran (47.0%), as discussed under Limitations. **Construct validity:** The tool-category taxonomy was grounded in a literature and market review of AI tools in design practice before survey deployment. Three subject-matter experts also reviewed the questionnaire for face and content validity. **Face validity** was assessed qualitatively through expert judgment on item clarity and relevance. **Content validity** was assessed with the Content Validity Index (Davis, 1992), computed from the three experts' relevance ratings on a 4-point scale. Item-level CVI was 1.00 for all items. Scale-level values were S-CVI/Ave = 1.00 and S-CVI/UA = 1.00, and both exceed standard thresholds (Polit & Beck, 2006). Participants reported on their most recently completed project, which anchored responses in a specific instance and reduced recall error. Recency (Q9) skewed toward very recent projects (Table 3). Projects completed today ($n = 197$, 44.5%) or within the past week ($n = 138$, 31.2%) made up most responses. **Data collection reliability:** Google Forms allowed one response per signed-in Google account and required every shown item to be answered before submission. This setting prevented duplicate and incomplete responses. **Measurement coverage and categorization:** No large-scale account of designers' reported AI tool use existed before this study, so the preset tool list was provisional. An open-text option was added to cover tools outside that list. Tool identification coverage was then checked in each phase to support the tool-count analyses, with results in Appendix Table 4. Two items measured phase use, and they did not always agree (Table 2). A binary indicator showed whether each phase was checked in the general item Q12, "In which of the following steps do you use AI?". McNemar's test compared it with the Yes or No phase screen (McNemar, 1947). The two items differed in every phase (Discover $\chi^2(1) = 32.58$, Define $\chi^2(1) = 24.77$, Develop $\chi^2(1) = 5.70$, Deliver $\chi^2(1) = 46.80$). All p values were below 0.001 except Develop ($p = 0.017$). In each phase, more participants confirmed use on the phase screen without checking it in Q12 than the reverse. Of the 93 distinct entries reviewed, 20 were excluded as not valid, identifiable AI tools. Some were ordinary design or office software with no AI feature. Others were ambiguous references that could not be matched to a specific product. The remaining 73 AI tools were classified. Agreement between raters (number of raters = 2) was assessed with Cohen's kappa, computed separately for Output and Function. Both dimensions showed almost perfect agreement (Output $\kappa = 1.00$,

Function κ = 0.937, n = 93; Landis & Koch, (1977). Disagreements were resolved through discussion until consensus was reached.

**Table 2: Recognition Effect between General and Phase-Specific AI-Use Items, by Phase**

| Phase | Mentioned in the general item Q12 | Confirmed "Yes" on phase-specific screen | Screen "Yes," not mentioned in Q12 | Mentioned in Q12, screen "No" |
|---|---|---|---|---|
| **Discover** | 265 | 318 | 68 | 15 |
| **Define** | 241 | 290 | 71 | 22 |
| **Develop** | 206 | 231 | 63 | 38 |
| **Deliver** | 107 | 182 | 96 | 21 |

***Note: Base is the 359 participants who answered "Yes" to the general screening question (Q10). Q12 is the general multi-select item on design steps. "Screen" is the Yes or No question that opened each phase section. McNemar's test with continuity correction compares the two discordant columns.***

## RESULTS

The final sample comprised 443 designers from at least 43 countries and regions (Table 3, Appendix Table 2). Participants covered five design roles plus an open "Other" category. Product and Industrial Designers were the largest groups at 26.9% each. UX/UI Designers followed (22.3%), then Graphic Designers (9.0%), Other (7.9%) and Design Students (7.0%). Respondents skewed early-career, with 60.3% aged 20 to 29 and 42.0% having fewer than four years of experience. Most held a Bachelor's (48.3%) or Master's degree (40.9%). Project type was mostly Digital (49.2%), followed by Hybrid (30.9%) and Physical (19.9%). Gender was not answered by 13.8% of participants, because the item was added partway through data collection. The sample was also geographically concentrated. Iran accounted for 47.0%, followed by Italy (7.9%) and India (7.7%). Another 2.5% preferred not to give their country. Results therefore describe this sample and should not be read as broadly cross-national.

**Table 3: Demographic Information (N = 443)**

| Variable | Category | n | % |
|---|---|---|---|
| **Age** | Under 20 | 7 | 1.6 |
| | 20 – 29 | 267 | 60.3 |
| | 30 – 39 | 133 | 30.0 |
| | 40 – 49 | 30 | 6.8 |
| | 50 – 59 | 4 | 0.9 |
| | 60 and above | 1 | 0.2 |
| | Prefer not to say | 1 | 0.2 |
| **Gender** | Male | 207 | 46.7 |
| | Female | 170 | 38.4 |
| | Non-binary | 3 | 0.7 |
| | Prefer not to say | 2 | 0.5 |
| | Not answered | 61 | 13.8 |
| **Education** | High school | 18 | 4.1 |
| | Bachelor's | 214 | 48.3 |
| | Master's | 181 | 40.9 |
| | Doctorate | 25 | 5.6 |
| | Other | 5 | 1.1 |
| **Experience** | < 1 year | 34 | 7.7 |
| | 1 – 3 years | 152 | 34.3 |
| | 4 – 6 years | 131 | 29.6 |
| | 7 + years | 126 | 28.4 |
| **Project Type** | Digital | 218 | 49.2 |
| | Physical | 88 | 19.9 |
| | Hybrid | 137 | 30.9 |
| **Role** | Product Designer | 119 | 26.9 |
| | Industrial Designer | 119 | 26.9 |
| | UX / UI Designer | 99 | 22.3 |
| | Graphic Designer | 40 | 9.0 |
| | Other | 35 | 7.9 |
| | Design Student | 31 | 7.0 |
| **Most Recent Project** | Today | 197 | 44.5 |
| | In the past week | 138 | 31.2 |
| | In the past month | 54 | 12.2 |
| | In the past 3 months | 31 | 7.0 |
| | More than 3 months | 23 | 5.2 |

***Note: "Not answered" for Gender reflects participants who completed the survey before the gender item was added partway through data collection. Percentages are of N = 443.***

## *AI Adoption across the Double Diamond*

Two measures of AI use existed (Table 1), a general screening question (Q10) and confirmed use within each phase section (Q13, Q23, Q33 and Q43). Six participants answered "Yes" to Q10 but confirmed no phase use. Adoption is therefore defined as confirmed use in at least one of the four phases, the more conservative and behavior-based measure. By this definition, 79.7% of the 443 designers qualified as adopters, and 20.3% showed no confirmed AI use in any phase. Adoption alone does not show how much of the workflow that use covers. Adopters used AI in a mean of 2.89 of the four phases (Table 4). Most used it in part of their process. Only 35.1% used AI in all four phases. The rest used it in two or three phases (55.8%) or in a single phase (9.1%).

**Table 4: Workflow Breadth among Confirmed Adopters (n = 353)**

| Breadth level | Adopters n | Adopters % | AI adoptions in Phases Combination | Breadth-level adopters n | Breadth-level adopters % |
|---|---|---|---|---|---|
| **1 phase** | 32 | 9.1 | **Discover only** | **14** | **43.8** |
| | | | Define only | 9 | 28.1 |
| | | | Develop only | 7 | 21.9 |
| | | | Deliver only | 2 | 6.3 |
| **2 phases** | 98 | 27.8 | **Discover + Define** | **58** | **59.2** |
| | | | Discover + Develop | 21 | 21.4 |
| | | | Discover + Deliver | 6 | 6.1 |
| | | | Develop + Deliver | 5 | 5.1 |
| | | | Define + Develop | 5 | 5.1 |
| | | | Define + Deliver | 3 | 3.1 |
| **3 phases** | 99 | 28.0 | **Discover + Define + Develop** | **57** | **57.6** |
| | | | Discover + Define + Deliver | 30 | 30.3 |
| | | | Discover + Develop + Deliver | 8 | 8.1 |
| | | | Define + Develop + Deliver | 4 | 4.0 |
| **4 phases** | 124 | 35.1 | **Discover + Define + Develop + Deliver** | **124** | **100.0** |

***Note: Breadth is the number of phases with confirmed AI use. Within each breadth level, rows list every phase combination used. Percentages in the last column are of that breadth level. Mean breadth = 2.89 (SD = 0.99).***

Partial adopters kept phases in workflow order. At every breadth level, the most common combination was the first phases of the workflow, added one at a time. Among one-phase adopters, Discover alone was the largest group (43.8%). **Discover + Define** led at breadth 2 (59.2%), and **Discover + Define + Develop** led at breadth 3 (57.6%). Each leading combination was the previous one plus the next phase (Table 4). Partial adoption therefore looked less like random use across the process and more like accumulation along the workflow. Use grew one phase at a time until it stopped or reached all four.

General frequency of use (Q11, n = 359) skewed toward habitual use. Most participants used AI daily (47.9%) or a few times a week (36.8%). Use once a month or less was rare (0.8%). Among adopters (n = 353), frequency and breadth were positively associated (Spearman's $\rho = 0.325$, $p < 0.001$). Daily users worked with AI in a mean of 3.21 phases, against 2.33 for those using it a few times a month (Figure 2). More frequent use went with wider workflow coverage.

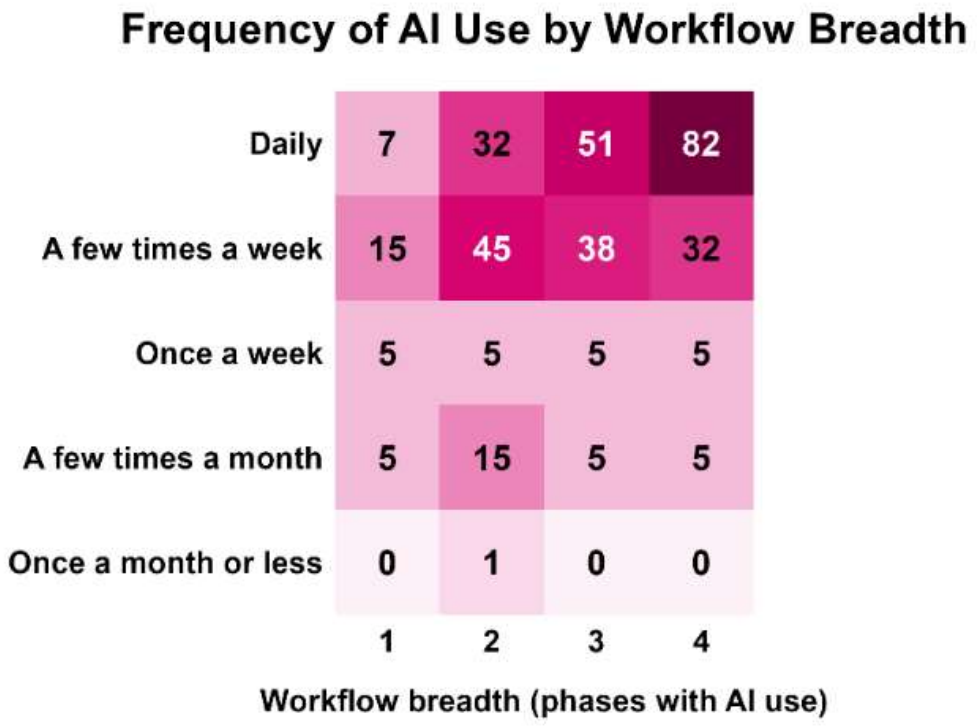

***Note: Cells show the number of adopters (n = 353) at each combination of general frequency (Q11) and workflow breadth. Breadth is the number of phases with confirmed AI use, from 1 to 4.***

**Figure 2: Frequency of AI Use by Workflow Breadth**

## *AI Tool Category across the Design Process*

Across the study, 73 distinct AI tools were identified, as described under AI Tool Classification. Table 5 gives the number of tools named, adopters and total tool picks in each phase. Distinct tools ranged from 32 in Deliver to 47 in Develop. Adopters picked a mean of 1.80 to 2.09 tools per phase.

Treating AI tool use as one behavior hides two choices a designer makes (Q. Lu et al., 2024). The first is which output the designer needs and another is the role the tool plays, either as a standalone generative tool or as an AI feature inside software the designer already uses.

**Table 5: Tools Named, Adopters and Tool Picks, by Phase**

| Phase | Adopters<br>n | AI-Tools in phase<br>n | AI-Tools total picks in phase<br>n | Picks per person<br>M |
|---|---|---|---|---|
| **Discover** | 318 | 40 | 664 | 2.09 |
| **Define** | 290 | 33 | 522 | 1.80 |
| **Develop** | 231 | 47 | 462 | 2.00 |
| **Deliver** | 182 | 32 | 344 | 1.89 |

**Output** refers to the primary modality a tool produces, from Text and Image to Video, 3D Geometry and Agentic. Audio is a recognized output category in the wider literature (Gozalo-Brizuela & Garrido-Merchán, 2023) but no participant reported a tool whose primary output was audio, so the category does not appear here. Voice is a medium for reading text in tools like ChatGPT, not a native voice-generating tool, and was coded as part of that tool's Text output rather than a distinct category (Appendix Table 3, Table 6). Figure 3 shows output modality by phase. **Text** led in Discover and Define, and **Image** led in Develop (72.1%). Deliver combined **Text** with **Image**. **Video** was absent in Define but grew through Develop and Deliver and exceeded **3D Geometry** in both.

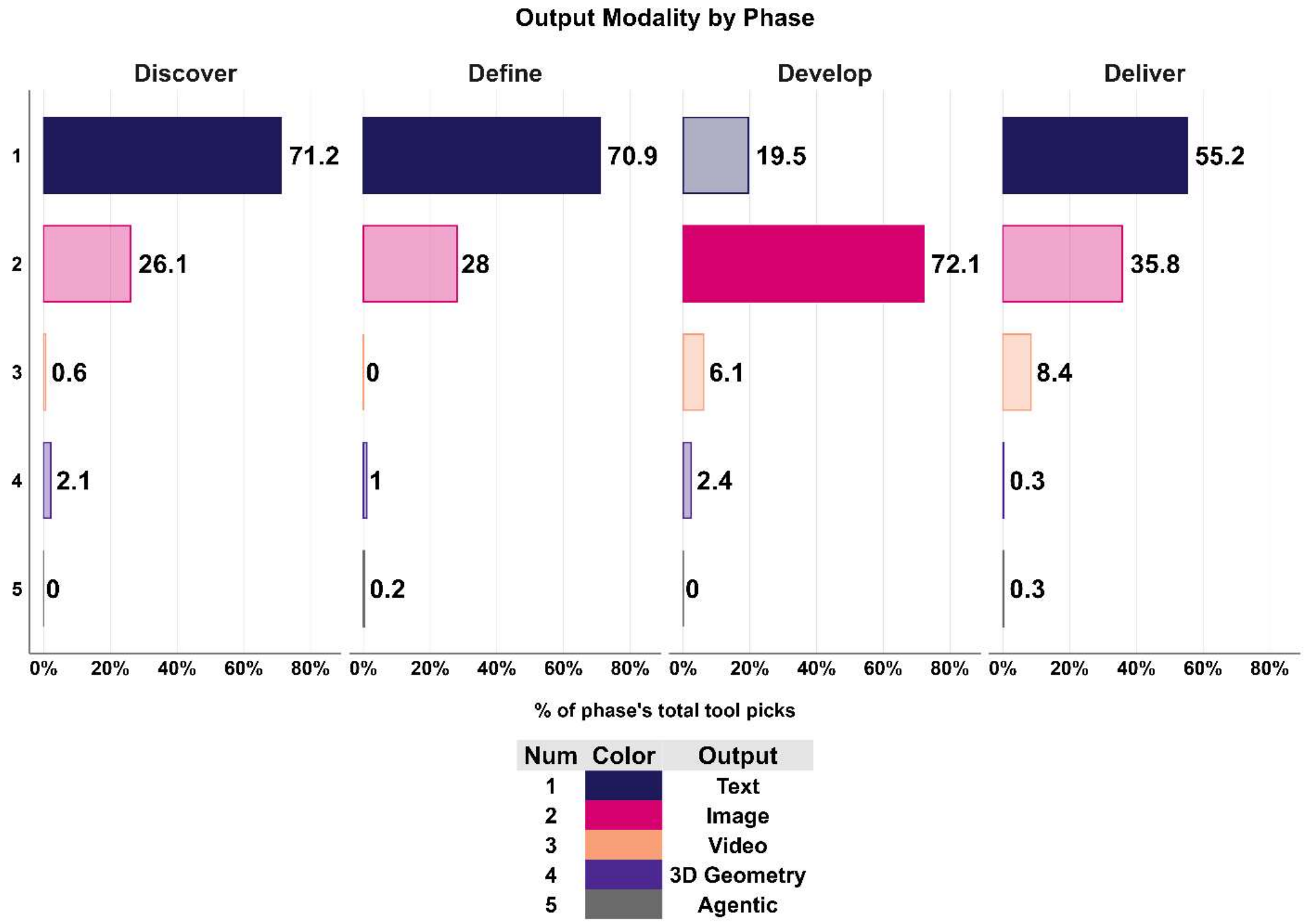

***Note: Bars show each output's share of that phase's total tool picks (Table 6). The most used output in each phase is shown in a darker color. Values are percentages.***

**Figure 3: Output Modality by Phase**

**Function** is the tool's role within its output. The scheme follows recent multi-dimensional approaches to generative AI systems (Shi et al., 2024; Doshi & Moore, 2026). Text had three functions: **Conversational & General-Purpose** tools are open-ended assistants such as ChatGPT and Gemini. **Research & Knowledge Synthesis** tools find and summarize information, as Perplexity AI and NotebookLM do. **Code Generation & Development** tools generate or scaffold code, as in Figma Make and Bolt.new. Image had two functions: **Design Software, Embedded AI** covers AI features inside existing design tools, like Figma AI and Adobe Photoshop. **AI-Native Image Generation** covers standalone prompt-to-image tools like Midjourney and Adobe Firefly. Video showed the same split: **AI-Native Video Generation** includes Runway ML, and **Video Editing Software, Embedded AI** includes Descript and Screen Studio. In 3D Geometry, Autodesk Fusion 360 fell under **CAD Platform, Embedded** AI and Tripo AI under **AI-Native**

**3D Generation**. Agentic tools form one function, **Multi-step Autonomous Execution**, which carries out tasks with little human direction. Manus AI and Marvin are its two tools. Figure 4 shows this finer level. Appendix Table 3 gives the tool-level data.

**Discover** is largely research and brainstorming. Within Text, it relied mainly on Conversational tools (Table 6). ChatGPT alone accounted for 45.0% of picks, above Research & Knowledge Synthesis tools like Perplexity AI. Designers may default to general-purpose chat for early sense-making. Within Image, Embedded Design Software features led by Figma AI outpaced AI-Native Image Generation. This result suggests a preference for AI inside existing tools over standalone generators (Appendix Table 3).

**Table 6: Nests Function within Each Output by Phase**

| Output | Function | Discover | | Define | | Develop | | Deliver | |
|---|---|---|---|---|---|---|---|---|---|
| | | n | % | n | % | n | % | n | % |
| **Text** | Conversational & General-Purpose | 384 | 57.8 | 322 | 61.7 | 60 | 13.0 | 178 | 51.7 |
| | Research & Knowledge Synthesis | 75 | 11.3 | 44 | 8.4 | 3 | 0.6 | 3 | 0.9 |
| | Code Generation & Development | 14 | 2.1 | 4 | 0.8 | 27 | 5.8 | 9 | 2.6 |
| **Image** | Design Software, Embedded AI | 134 | 20.2 | 131 | 25.1 | 95 | 20.6 | 114 | 33.1 |
| | AI-Native Image Generation | 39 | 5.9 | 15 | 2.9 | 238 | 51.5 | 9 | 2.6 |
| **Video** | AI-Native Video Generation | 4 | 0.6 | 0 | 0.0 | 28 | 6.1 | 15 | 4.4 |
| | Video Editing Software, Embedded AI | 0 | 0.0 | 0 | 0.0 | 0 | 0.0 | 14 | 4.1 |
| **3D Geometry** | CAD Platform, Embedded AI | 13 | 2.0 | 5 | 1.0 | 11 | 2.4 | 1 | 0.3 |
| | AI-Native 3D Generation | 1 | 0.2 | 0 | 0.0 | 0 | 0.0 | 0 | 0.0 |
| **Agentic** | Multi-step Autonomous Execution | 0 | 0.0 | 1 | 0.2 | 0 | 0.0 | 1 | 0.3 |
| **Total** | | **664** | | **522** | | **462** | | **344** | |

***Note: n = combined tool selections within that Output and Function pair, for that phase. % = n ÷ that phase's total tool selections (664, 522, 462, 344), calculated as in Appendix Table 3 but aggregated to the category level.***

In **Define**, research results are synthesized into a problem statement. The functional preference matched Discover, and Conversational tools rose further, still led by ChatGPT. Embedded tools shifted from Figma AI to FigJam AI. This shift follows the move from open exploration to structured clustering and diagramming (Table 6, Appendix Table 3).

In **Develop**, solutions are generated and prototyped, and here function inverted. AI-Native Image Generation, led by Midjourney, overtook Conversational tools (51.5% vs 13.0% of picks). This change is the clearest shift in the workflow, from general-purpose assistance to purpose-built generation. Code Generation tools also grew here, in line with Develop's broader move toward specialized, output-specific functions (Appendix Table 3).

In **Deliver**, solutions are tested and refined. Conversational tools led again, below their share in Discover and Define. Embedded Design Software, led by Adobe Photoshop, reached its highest share of any phase. Deliver was the only phase that used Embedded Video Editing tools (Table 6, Appendix Table 3).

**Develop** also had the largest number of distinct tools (Table 5). This count may reflect the many AI-Native Image and Video products on the market. By contrast, Discover and Define centered on one stable tool, ChatGPT. Develop is also where the designer's role may change most visibly, from directing a conversation to curating machine output. Studies of designer and AI co-creation (Guo et al., 2023) and of avatar design (He et al., 2024) describe this change. Only 9 of the 73 tools (12.3%) were Conversational & General-Purpose. Yet these few tools received the majority of picks in three of the four phases: use concentrated on a small set of broadly capable tools. In variety, however, the tool ecosystem was mostly narrow and task-specific.

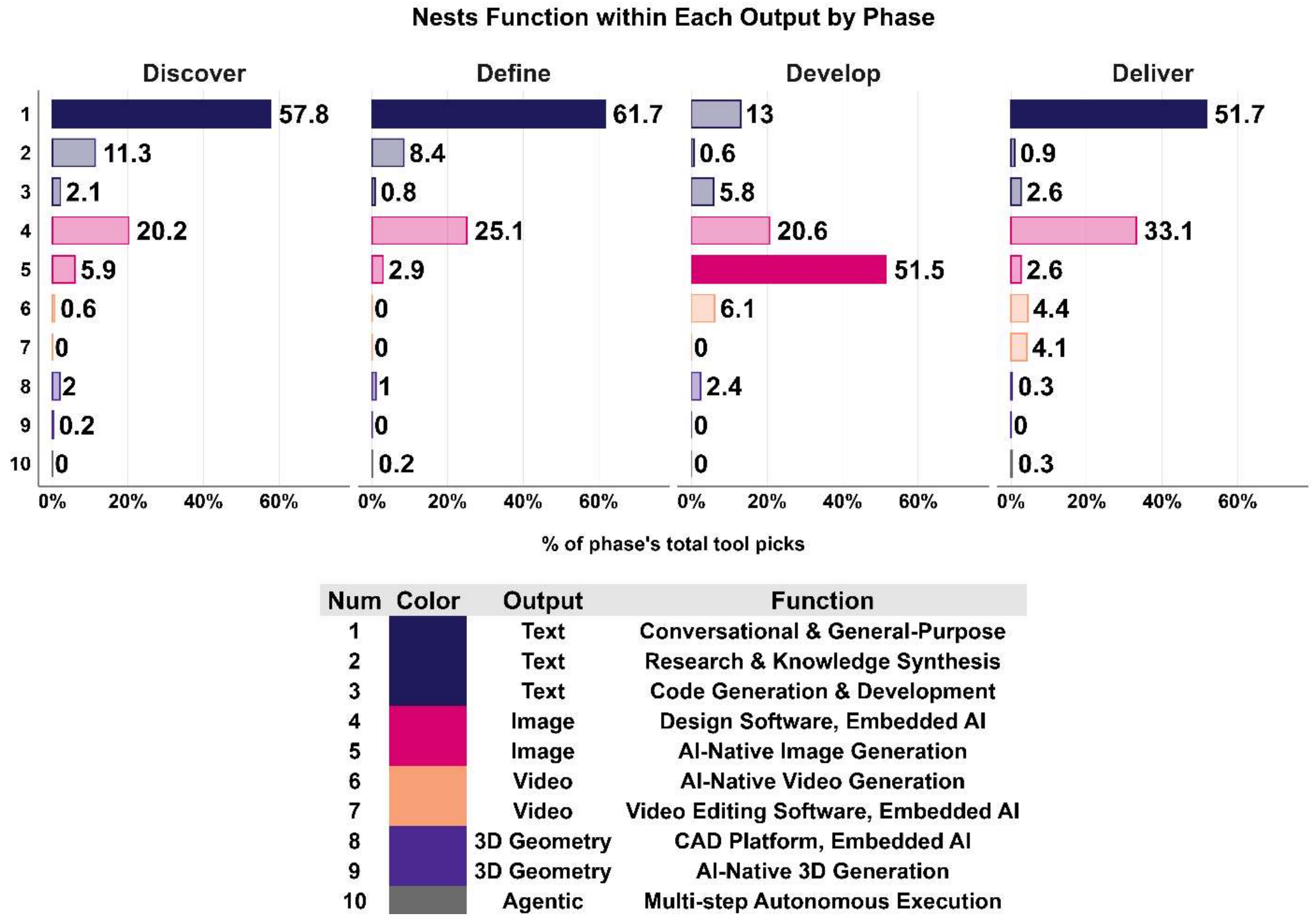

***Note: Bars show each of the 10 function-nests as a share of that phase's total tool picks (Table 6). The most used function in each phase is shown in a darker color. Values are percentages.***

**Figure 4: Nests Function within Each Output by Phase**

Grouping tools by mode instead of output shows a second pattern (Table 7, Figure 5). **AI-Native Generation tools** were rare in Discover and Define, rose to 57.6% of picks in Develop and fell again in Deliver. **Embedded AI tools** stayed stable through the first three phases and peaked in Deliver at 37.5%. The two modes peak in different phases. AI-Native tools serve the creation of new material, mostly in Develop. Embedded AI is used more evenly and leads over AI-Native tools when work is refined and finalized in Deliver.

**Table 7: AI-Native Tools vs. Embedded AI Tools Use, by Phase**

| Phase | AI-Native Tools | | Embedded AI Tools | | Total-picks |
|---|---|---|---|---|---|
| | n | % | n | % | n |
| **Discover** | 44 | 6.6 | 147 | 22.1 | 664 |
| **Define** | 15 | 2.9 | 136 | 26.1 | 522 |
| **Develop** | 266 | 57.6 | 106 | 22.9 | 462 |
| **Deliver** | 24 | 7.0 | 129 | 37.5 | 344 |

***Note: % = share of that phase's total tool picks. AI-Native Tools = AI-Native Image + Video + 3D Generation combined. Embedded AI Tools = Design Software + Video Editing Software + CAD Platform Embedded AI combined. Shares do not sum to 100% because Text and Agentic tools are excluded.***

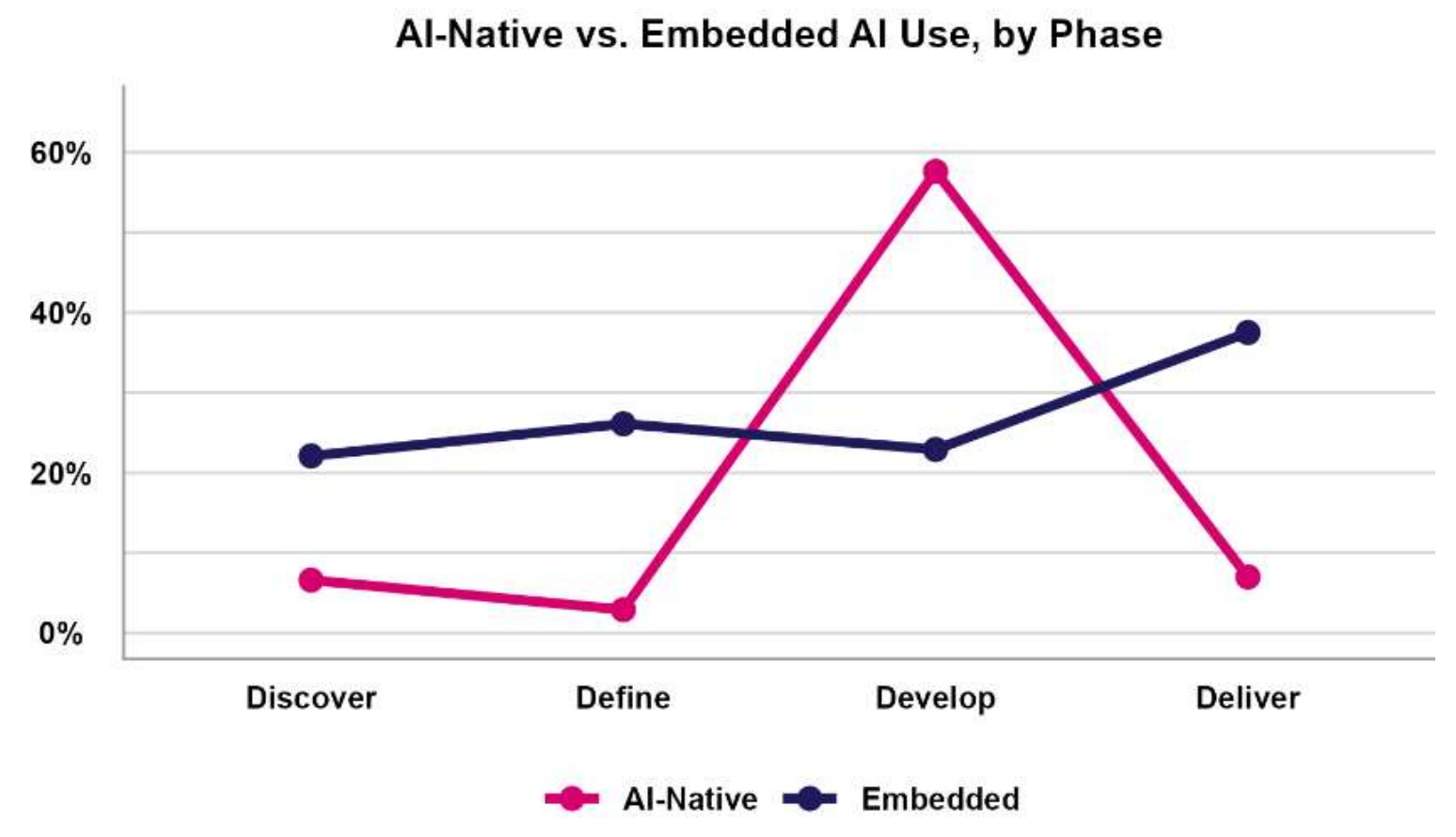

***Note: Lines show the shares in Table 7. Totals per phase are 664, 522, 462 and 344 picks (Table 5).***

**Figure 5: AI-Native Tools vs. Embedded AI Tools Use, by Phase**

## *AI Use Intensity across Workflow*

Before turning to intensity, tool identification coverage was checked to confirm reported AI use could reliably be linked to specific tools. In every phase, at least 93.1% of reported use was linked to an identifiable tool (Appendix Table 4).

Two levels of analysis are reported, because the branching design skipped non-adopters in each phase. The population and adopter views tell different stories. At the population level (N = 443, non-adopters coded as zero), tool count differed across phases (Friedman $\chi^2(3) = 133.87$, $p < 0.001$). It was highest in Discover and lowest in Deliver (Friedman, 1937). The effect size was small (Kendall's W = 0.101), so phase explains only a limited share of the variance. Holm-corrected Wilcoxon signed-rank tests showed that all six pairs of phases differed (all Holm-adjusted $p < 0.05$). Effect sizes ranged from r = 0.105 for Define vs Develop to r = 0.455 for Discover vs Deliver. Among adopters, intensity did not decline steadily. It dropped from Discover to Define, rose again in Develop and eased slightly in Deliver (Figure 6, Table 8). Counts ranged from 0 to a maximum of 10 tools, reached in Define. This adopter pattern gives the more informative account of depth once someone is already using AI. The population-level decline reflects fewer designers using AI at all in later phases, from 71.8% in Discover to 41.1% in Deliver. Adopters who stayed did not scale back their tool use. A few adopters in each phase named no identifiable tool, which explains the minimum of 0 (Table 8, Appendix Table 4).

**Table 8: AI-Use Intensity across the Workflow**

| Phase | Population | Population | | Adopters | | Adopters | | Adopters | Adopters | Adopters | |
|---|---|---|---|---|---|---|---|---|---|---|---|
| | N | M | SD | n | % | M | SD | Median | IQR | Min | Max |
| **Discover** | 443 | 1.50 | 1.43 | 318 | 71.8 | 2.09 | 1.27 | 2 | 1 – 3 | 0 | 9 |
| **Define** | 443 | 1.18 | 1.26 | 290 | 65.5 | 1.80 | 1.15 | 1 | 1 – 2 | 0 | 10 |
| **Develop** | 443 | 1.04 | 1.43 | 231 | 52.1 | 2.00 | 1.42 | 2 | 1 – 3 | 0 | 7 |
| **Deliver** | 443 | 0.78 | 1.17 | 182 | 41.1 | 1.89 | 1.11 | 2 | 1 – 2 | 0 | 7 |

***Note: The minimum of 0 comes from a few adopters who reported AI use in a phase but named no identifiable tool (4, 2, 16 and 4 in Discover, Define, Develop and Deliver, Appendix Table 4). This group is too small to shift the 25th percentile, so the IQR still starts at 1.***

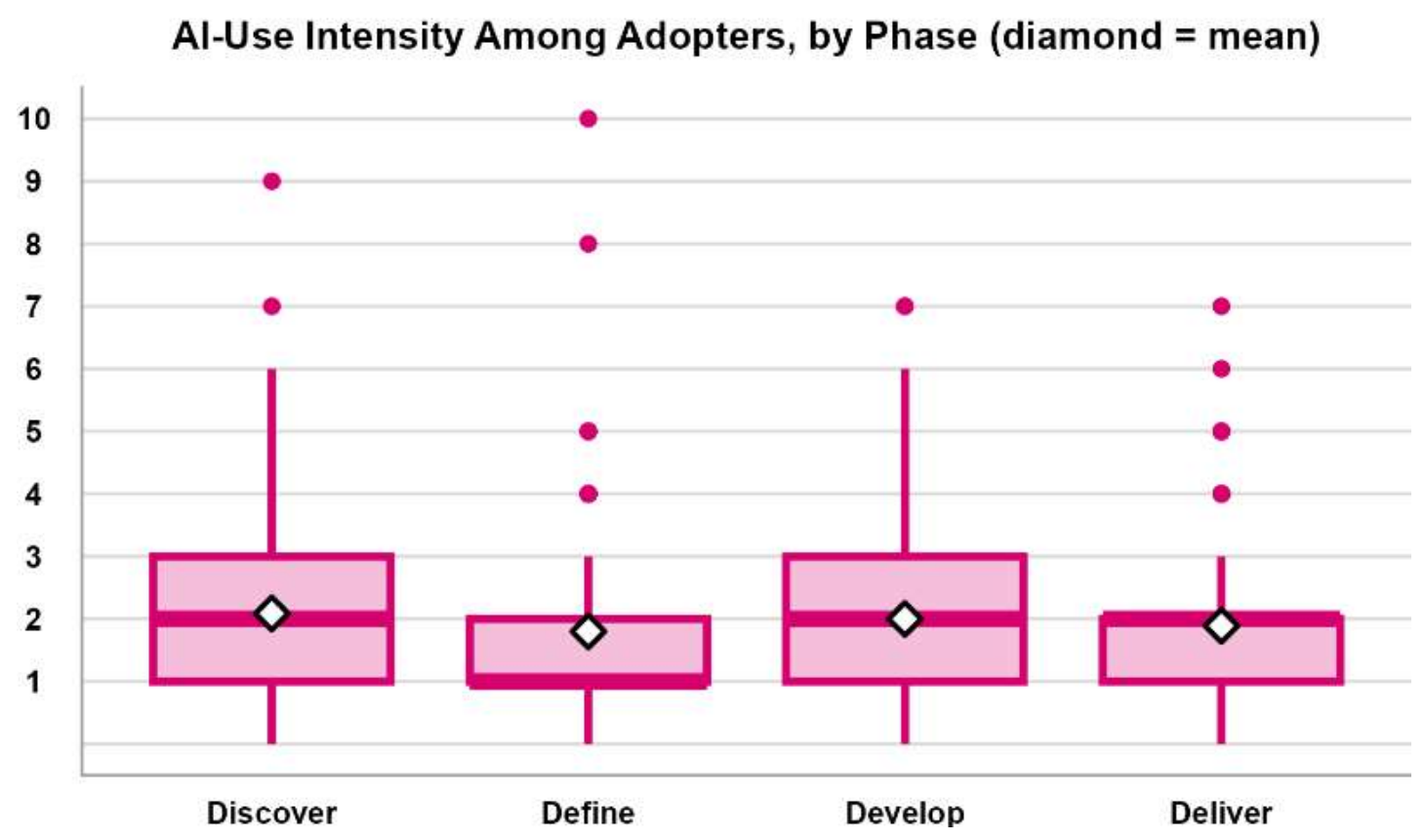

***Note: Boxplot shows tool-count distribution among confirmed adopters only (n = 318, 290, 231, 182 for Discover, Define, Develop, Deliver respectively; see Table 8).***

**Figure 6: AI-Use Intensity among Adopters (n = 318, 290, 231, 182), by Phase**

## *Barriers within AI Adopters and Non-adopters*

RQ4 asked what barriers designers report regarding AI use, and how these differ between adopters and non-adopters. The two groups reached Q62 at different points in the survey: **non-adopters** were routed there immediately after answering 'No' (Q10), while **adopters** answered only after completing all four phase-specific sections (Table 1). This asymmetry in question position is a relevant contextual factor when interpreting differences in endorsement between the two groups, alongside the differing group sizes (359 vs. 84). Because this was a multi select item, each of the ten barrier categories was tested separately as a 2×2 chi-square comparison, Holm-corrected, with each category calculated as a percentage of its own group (adopters n=359; non-adopters n=84); percentages don't sum to 100% within either group, since one person could be counted in multiple rows. Effect size was reported as φ (Pearson, 1900; Holm, 1979).

**Where the groups differed:** only one barrier survived Holm correction, **low perceived usefulness**, endorsed by 40.5% of non-adopters vs. 16.4% of adopters ($\chi^2$=22.29, p_Holm<0.001, φ=0.224, small-to-medium). Two others were nominally significant but did not survive correction: cost (37.6% vs. 23.8%) and job security (13.4% vs. 4.8%). Both were endorsed more by adopters than by non-adopters, the opposite direction to usefulness, but these differences should not be interpreted. **Accuracy/reliability** was the second most-endorsed barrier overall and showed no statistically significant group difference (32.9% vs. 36.9%), so it was similarly endorsed in both groups, unlike usefulness (Table 9; Figure 7).

**Where the groups agreed:** the remaining seven barriers (lack of awareness, ease of use, accuracy/reliability, ethical concerns, organizational constraints, workflow incompatibility, and "Other") showed no statistically significant difference between adopters and non-adopters (all Holm-adjusted p = 1.000; Table 9; Figure 7). These barriers were endorsed at similar rates in both groups in this sample, but non-significance does not establish equivalence (Table 9; Figure 7).

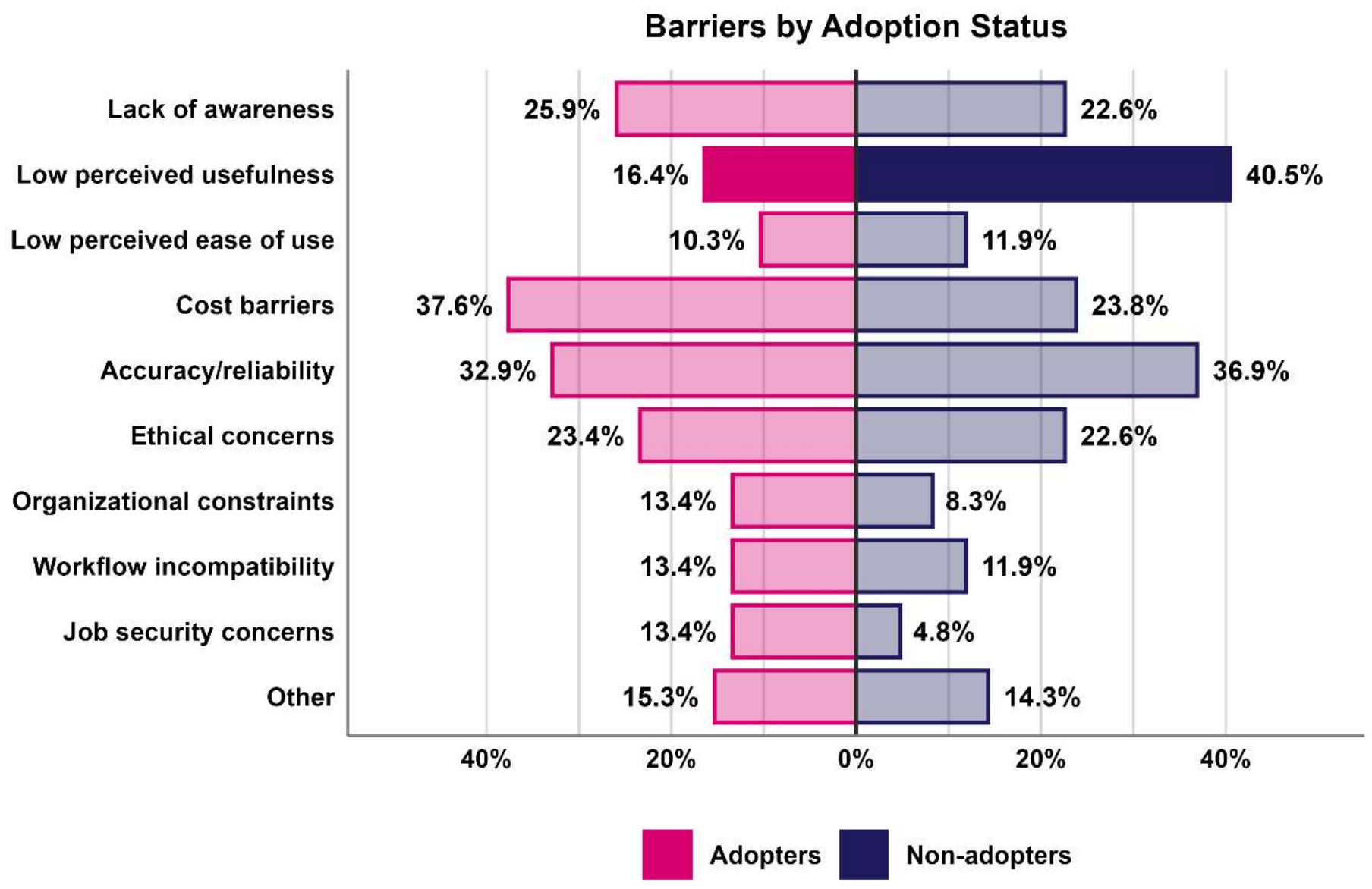

***Note: Bars show each barrier's endorsement rate within its own group (adopters n = 359, non-adopters n = 84). Solid bars mark the barrier significant after Holm correction.***

**Figure 7: Barriers by Adoption**

**Table 9: Barriers by Adoption Status**

| Barrier | Adopters | | Non-adopters | | p(Holm) | φ |
|---|---|---|---|---|---|---|
| | n | % | n | % | | |
| **Lack of awareness** | 93 | 25.9 | 19 | 22.6 | 1.000 | 0.023 |
| **Low perceived usefulness** | **59** | **16.4** | **34** | **40.5** | **<0.001**** | **0.224** |
| **Low perceived ease of use** | 37 | 10.3 | 10 | 11.9 | 1.000 | 0.011 |
| **Cost barriers** | 135 | 37.6 | 20 | 23.8 | 0.215 | 0.107 |
| **Accuracy/reliability** | 118 | 32.9 | 31 | 36.9 | 1.000 | 0.027 |
| **Ethical concerns** | 84 | 23.4 | 19 | 22.6 | 1.000 | 0.0004 |
| **Organizational constraints** | 48 | 13.4 | 7 | 8.3 | 1.000 | 0.051 |
| **Workflow incompatibility** | 48 | 13.4 | 10 | 11.9 | 1.000 | 0.009 |
| **Job security concerns** | 48 | 13.4 | 4 | 4.8 | 0.348 | 0.096 |
| **Other** | 55 | 15.3 | 12 | 14.3 | 1.000 | 0.003 |

***Note: ** p(Holm) < 0.001, the only barrier significant after Holm correction. Chi-square tests used Yates correction.***

# DISCUSSION

Three patterns in this dataset run against what the literature has assumed about where AI concentrates in design work. Lee et al. (2025) found AI design support systems built mostly for Develop and Deliver, with early-stage Discover and Define comparatively neglected. Our data show the reverse: Discover had the highest adoption of any phase, led by Conversational, General-Purpose tools. This result is closer to the finding of Chen et al. (2025) that generative AI assists problem definition and ideation while evaluation stays human-led. It fits less well with a picture of purpose-built early-stage tools. The tool doing this early work was not designed for design.

Define shows the same pattern. It remains one of the workflow's two most AI-saturated phases and shows no gap. Zhou and Chen (2025) found Define supported by LLM functions just as developed as the other three phases. Our adoption data agree.

This concentration is cumulative. At every breadth level, the leading combination adds the next phase in order instead of skipping one (Table 4). This staged, additive pattern resembles diffusion of innovation accounts of technology adoption (Rogers, 2003). Here it appears within one workflow instead of across a population over time. Frequent users spread their use across more phases, so habitual use and breadth go together.

Just 9 of the 73 tools identified are general-purpose. Yet they account for most picks in three of four phases. Fourteen tools appear in all four phases and form a stable core. Another 33 appear in one phase only. The broader synthesis of Mirzaei et al. (2026) found AI's role differentiated by stage instead of concentrated at either end. This dataset shows a sharper early-phase skew than that picture suggests.

Develop is the one phase where the data agree with the built tool field. Conversational use falls sharply here as AI-Native Image Generation leads. This pattern matches the account of Lee et al. (2025) of where the field has concentrated its tool-building. Develop is also where the designer's role may shift most visibly, from directing a conversation to curating machine output. Adoption drops here relative to the first two phases. Intensity among engaged users rises again from Define. AI-Native and Embedded modes peak in different phases, Develop and Deliver. Two distinct kinds of human-AI collaboration may be at work, instead of one behavior at different intensities.

Deliver shows the lowest adoption of any phase. Built systems would least predict this pattern, given how much late-stage tooling exists. Among image tools, designers in Deliver chose Embedded Design Software features over standalone generative systems. Süner-Pla-Cerdà et al. (2026) found that industrial designers frame adoption around a tension between automation and designerly autonomy. That tension fits Deliver, where a finalized decision is hardest to revisit. The intensity data point the same way. At the population level, the decline reflects fewer designers using AI in later phases while engaged users did not scale back. Each phase seems to pose a fresh adoption decision.

Comparing adopters and non-adopters on the same barrier measure reveals a distinction a single-group study could not. The two groups differed on one dimension only: non-adopters judged available tools as not useful enough for their work. Unfamiliarity, ethics and organizational restriction did not separate them. This result supports the persona-based account of Fang, Guo, et al. (2026) of non-adoption as differentiated by usefulness. It also sharpens that account, since usefulness alone separates the groups. The same judgment may explain where adopters concentrate their use and why non-adopters stay out.

These patterns are descriptive and do not show mechanisms. Diffusion of innovation or technology acceptance models could test why usefulness judgments, more than familiarity or ease of use, appear linked to both adoption and non-adoption. Such a test lies beyond this study's descriptive scope.

### *Limitations*

This study has several limitations. Recruitment relied on unpaid, non-probability sampling through LinkedIn outreach, design-focused Facebook and Telegram groups and snowball distribution. Over about six months, it produced a sample concentrated in Iran (47.0%), which limits generalization to a wider global population. The branching design also meant that phase-specific samples declined across the workflow (Discover $n = 318$ to Deliver $n = 182$). Cross-phase comparisons are weaker for small subgroups. All measures were self-reported, though 75.6% of reference projects were completed today or within the past week. Recall bias was likely small. The cross-sectional design rules out causal inference, so associations between phase, tool category and adoption should be read descriptively.

The instrument itself carries further limits. It was administered in English, which was not the first language of most respondents. Non-native speakers were told they could use a translation tool such as Google Translate, but this remedy was informal and its use was not tracked. The preset tool lists may have biased reported use toward the tools offered. An open-text option was meant to offset this bias. Every open-text entry still required manual identification and classification, which added researcher judgment absent from preset selections. The general multi-select item (Q12) also showed a significant recognition gap against each phase's screening question (McNemar's $\chi^2 = 5.70$ to $46.80$, all $p < 0.05$). Participants recognized phase-specific AI use more readily in the concrete phase items than in the abstract general item. Q12 alone would therefore have undercounted adoption relative to the phase screens used here. Custom items for tool selection and barriers were reviewed for face and content validity but not validated further. Only the NASA-TLX block is a validated measure. Ethics, risk-perception and KPI items were collected in the same instrument but are left to companion analyses.

### *Suggestions for Future Research*

Several directions follow. Longitudinal work could test whether new purpose-built early-stage tools shift adoption toward the later-stage concentration that current systems were designed for. It could also show whether Conversational & General-Purpose AI keeps its lead regardless. Qualitative follow-up could explain what this descriptive account cannot. One question is why partial adopters keep Discover and Define but drop Deliver. Another is why most designers converge on a small shared set of tools while a minority try idiosyncratic alternatives. Interviews could also ask why designers choose specific tools and what each phase demands from an AI collaborator: conversational support, generative output or an embedded feature. Role-level differences in adoption, frequency and breadth call for hypothesis-driven tests across the largest design roles. Such tests could check whether perceived usefulness, the one barrier separating adopters from non-adopters here, carries the same weight within each role. Other factors such as years of experience and project type may also shape adoption, and this dataset could test them. A mixed-effects or hurdle-model analysis of the same data would extend the intensity results into a fully inferential account.

## CONCLUSION

This study examined how 443 designers from at least 43 countries and regions reported using AI across the four phases of the Double Diamond. It drew on a taxonomy of 73 distinct tools classified by output and function. Adoption had several dimensions. It mattered whether designers adopted at all and how much of the workflow their use spanned. The kind of tool chosen at each stage also mattered, as did how intensely designers used it once engaged. A small core of tools recurred across all four phases. Most other tools appeared in only some phases. Depth of use shifted unevenly across the workflow and did not decline uniformly. The central result runs against expectation, since practice may be running ahead of tooling. Conversational & General-Purpose AI filled early-stage work where few purpose-built systems exist. AI-Native and Embedded AI use peaked in different phases, which points to two distinct modes of collaboration. Barriers were specific too. Usefulness was the one concern that separated adopters from non-adopters, while unfamiliarity and ethics did not. AI adoption in design is therefore better treated as several behaviors than as one behavior measured once. For designers, educators and industry, these phase and mode patterns offer a practical map of where AI currently earns its place and where human judgment is still preferred.

## *Acknowledgments*

The authors thank the 443 designers from at least 43 countries and regions who took part in this study. They also thank the design communities and networks that helped distribute the survey.

## *CRediT Author Statement*

Sepideh Tajarmakan: Conceptualization, Methodology, Investigation, Data curation, Formal analysis, Software, Visualization, Writing – original draft. Khashayar Hojjati Emami: Supervision, Validation, Writing – review & editing.

## *Consent and Ethical Considerations*

The questionnaire's first item (Q1) presented an informed consent statement. It read, "By continuing, you agree to participate in this anonymous research study on AI tools in design workflows. Your responses will be confidential and used only for academic purposes." Participants selected "Yes" or "No" to "I consent to participate", and only "Yes" respondents proceeded. Participation was voluntary, and participants could withdraw at any point before submission. No names, telephone numbers or other personal details were collected. An optional final item invited an email address for a possible follow-up interview. No interviews took place. All email addresses were removed before analysis, and the deposited files contain no personal data. The original export is held by the Department of Industrial Design, Iran University of Art, and is not shared. This department reviewed the study and confirmed in writing that it did not require an ethics approval code.

## *Declaration of Generative AI Use*

Grammarly was used for grammar and language editing. Claude (Anthropic) was used to refine figure design. All data analysis, statistical testing and interpretation were conducted by the authors, who reviewed and edited all text and take full responsibility for the content.

## Data Availability Statement

The analysis code (R), codebook and READ.md are available on Zenodo: https://doi.org/10.5281/zenodo.24239213. The tool taxonomy, cleaned datasets and validity coding files are deposited in a second Zenodo record: https://doi.org/10.5281/zenodo.24241427 currently under restricted access. Restricted access reflects the fact that this dataset underlies an ongoing paper series drawing on the same instrument, not yet complete. Access requests are reviewed by the corresponding author and granted case by case. These materials will be made openly available upon completion of the dissertation series. Running the code with these files reproduces every table and figure.

## Funding

This work received no external funding.

## Declaration of Interest Statement

The authors report there are no competing interests to declare.

# APPENDIX

## Appendix Table 1: Example of Binary Tool Coding (n = 3 Participants)

| Participant | Open-text response | ChatGPT | Grok | Gemini | Siemens NX | Altair Inspire |
|---|---|---|---|---|---|---|
| **1** | Chatgpt | 1 | 0 | 0 | 0 | 0 |
| **2** | Grok | 0 | 1 | 0 | 0 | 0 |
| **3** | Siemens NX, Altair Inspire | 0 | 0 | 0 | 1 | 1 |

***Note: Example of binary tool coding for three participants' open-text responses, out of 93 tool entries reviewed. A tool a participant named or selected is coded 1. All other tool columns, including unselected preset options, are coded 0. The same coding applies to preset and open-text tools in all four phases.***

## Appendix Table 2: Country or Region (N = 443)

| Country | n | % |
|---|---|---|
| Iran | 208 | 47.0 |
| Italy | 35 | 7.9 |
| India | 34 | 7.7 |
| Turkiye | 25 | 5.6 |
| United States | 19 | 4.3 |
| Canada | 12 | 2.7 |
| Germany | 12 | 2.7 |
| Sri Lanka | 8 | 1.8 |
| Pakistan | 7 | 1.6 |
| Spain | 7 | 1.6 |
| Australia | 6 | 1.4 |
| France | 5 | 1.1 |
| South Africa | 5 | 1.1 |
| Portugal | 4 | 0.9 |
| United Kingdom | 4 | 0.9 |
| Egypt | 4 | 0.9 |
| Argentina | 3 | 0.7 |
| Singapore | 3 | 0.7 |
| Bangladesh | 2 | 0.5 |
| Bosnia and Herzegovina | 2 | 0.5 |
| Nigeria | 2 | 0.5 |
| Sweden | 2 | 0.5 |
| Indonesia | 2 | 0.5 |
| Russia | 2 | 0.5 |
| Belgium | 1 | 0.2 |
| Japan | 1 | 0.2 |
| Taiwan | 1 | 0.2 |
| Vietnam | 1 | 0.2 |
| Finland | 1 | 0.2 |
| Jordan | 1 | 0.2 |
| Netherlands | 1 | 0.2 |
| New Zealand | 1 | 0.2 |
| Ethiopia | 1 | 0.2 |
| Hong Kong | 1 | 0.2 |
| Ireland | 1 | 0.2 |
| Kenya | 1 | 0.2 |
| Malaysia | 1 | 0.2 |
| Mexico | 1 | 0.2 |
| Peru | 1 | 0.2 |
| South Korea | 1 | 0.2 |
| Switzerland | 1 | 0.2 |
| Albania | 1 | 0.2 |
| Denmark | 1 | 0.2 |
| **Prefer not to say** | 11 | 2.5 |
| **Total** | **443** | **100%** |

## Appendix Table 3: Nests Function within Each Output by Phase, with AI Tool Names

| Output | Function | Tool | Discover | | Define | | Develop | | Deliver | |
|---|---|---|---|---|---|---|---|---|---|---|
| | | | n | % | n | % | n | % | n | % |
| Text | Code Generation & Development | Figma Make | 3 | 0.5 | 1 | 0.2 | 6 | 1.3 | 2 | 0.6 |
| | | Lovable | 3 | 0.5 | 1 | 0.2 | 4 | 0.9 | 1 | 0.3 |
| | | Google AI Studio | 2 | 0.3 | 1 | 0.2 | 3 | 0.6 | 2 | 0.6 |
| | | UX Pilot | 2 | 0.3 | 0 | 0.0 | 2 | 0.4 | 0 | 0.0 |
| | | Bolt.new | 1 | 0.2 | 0 | 0.0 | 2 | 0.4 | 1 | 0.3 |
| | | Cursor | 1 | 0.2 | 0 | 0.0 | 1 | 0.2 | 0 | 0.0 |
| | | Framer AI | 1 | 0.2 | 0 | 0.0 | 0 | 0.0 | 1 | 0.3 |
| | | Vertex AI | 1 | 0.2 | 1 | 0.2 | 0 | 0.0 | 0 | 0.0 |
| | | Codex | 0 | 0.0 | 0 | 0.0 | 2 | 0.4 | 0 | 0.0 |
| | | GitHub Copilot | 0 | 0.0 | 0 | 0.0 | 0 | 0.0 | 1 | 0.3 |
| | | Google Stitch | 0 | 0.0 | 0 | 0.0 | 2 | 0.4 | 0 | 0.0 |
| | | Replit | 0 | 0.0 | 0 | 0.0 | 2 | 0.4 | 0 | 0.0 |
| | | Rocket.new | 0 | 0.0 | 0 | 0.0 | 1 | 0.2 | 0 | 0.0 |
| | | V0 | 0 | 0.0 | 0 | 0.0 | 2 | 0.4 | 1 | 0.3 |
| | Conversational & General-Purpose | ChatGPT* | 299 | 45.0 | 273 | 52.3 | 29 | 6.3 | 155 | 45.1 |
| | | Gemini | 51 | 7.7 | 31 | 5.9 | 20 | 4.3 | 15 | 4.4 |
| | | Claude | 11 | 1.7 | 8 | 1.5 | 2 | 0.4 | 2 | 0.6 |
| | | Microsoft Copilot | 10 | 1.5 | 3 | 0.6 | 3 | 0.6 | 3 | 0.9 |
| | | Grok | 8 | 1.2 | 2 | 0.4 | 6 | 1.3 | 3 | 0.9 |
| | | DeepSeek | 3 | 0.5 | 2 | 0.4 | 0 | 0.0 | 0 | 0.0 |
| | | Kimi | 2 | 0.3 | 1 | 0.2 | 0 | 0.0 | 0 | 0.0 |
| | | Dia Browser | 0 | 0.0 | 1 | 0.2 | 0 | 0.0 | 0 | 0.0 |
| | | Qwen | 0 | 0.0 | 1 | 0.2 | 0 | 0.0 | 0 | 0.0 |
| | Research & Knowledge Synthesis | Perplexity AI* | 55 | 8.3 | 2 | 0.4 | 1 | 0.2 | 0 | 0.0 |
| | | Elicit* | 9 | 1.4 | 0 | 0.0 | 0 | 0.0 | 0 | 0.0 |
| | | NotebookLM | 6 | 0.9 | 8 | 1.5 | 2 | 0.4 | 2 | 0.6 |
| | | Dovetail | 2 | 0.3 | 1 | 0.2 | 0 | 0.0 | 0 | 0.0 |
| | | Notion AI* | 2 | 0.3 | 33 | 6.3 | 0 | 0.0 | 1 | 0.3 |
| | | Consensus | 1 | 0.2 | 0 | 0.0 | 0 | 0.0 | 0 | 0.0 |
| Image | AI-Native Image Generation | Vizcom | 15 | 2.3 | 4 | 0.8 | 11 | 2.4 | 1 | 0.3 |
| | | Midjourney* | 10 | 1.5 | 2 | 0.4 | 100 | 21.6 | 0 | 0.0 |
| | | Nano Banana | 7 | 1.1 | 3 | 0.6 | 10 | 2.2 | 2 | 0.6 |
| | | Krea | 3 | 0.5 | 2 | 0.4 | 2 | 0.4 | 2 | 0.6 |
| | | Leonardo AI | 3 | 0.5 | 1 | 0.2 | 3 | 0.6 | 0 | 0.0 |
| | | Visual Electric | 1 | 0.2 | 1 | 0.2 | 1 | 0.2 | 0 | 0.0 |
| | | Adobe Firefly* | 0 | 0.0 | 2 | 0.4 | 45 | 9.7 | 1 | 0.3 |
| | | ComfyUI | 0 | 0.0 | 0 | 0.0 | 1 | 0.2 | 0 | 0.0 |
| | | DALL·E* | 0 | 0.0 | 0 | 0.0 | 56 | 12.1 | 0 | 0.0 |
| | | Flora AI | 0 | 0.0 | 0 | 0.0 | 1 | 0.2 | 1 | 0.3 |
| | | FLUX.1 | 0 | 0.0 | 0 | 0.0 | 2 | 0.4 | 0 | 0.0 |
| | | Lovart | 0 | 0.0 | 0 | 0.0 | 1 | 0.2 | 0 | 0.0 |
| | | Lummi | 0 | 0.0 | 0 | 0.0 | 1 | 0.2 | 0 | 0.0 |
| | | **Microsoft designer** | 0 | 0.0 | 0 | 0.0 | 2 | 0.4 | 0 | 0.0 |
| | | **Recraft AI** | 0 | 0.0 | 0 | 0.0 | 1 | 0.2 | 0 | 0.0 |
| | | **Stable Diffusion** | 0 | 0.0 | 0 | 0.0 | 0 | 0.0 | 1 | 0.3 |
| | | **Topaz Labs** | 0 | 0.0 | 0 | 0.0 | 0 | 0.0 | 1 | 0.3 |
| | | **Whisk** | 0 | 0.0 | 0 | 0.0 | 1 | 0.2 | 0 | 0.0 |
| | Design Software, Embedded AI | **Figma AI*** | 72 | 10.8 | 0 | 0.0 | 88 | 19.0 | 3 | 0.9 |
| | | **Miro AI*** | 37 | 5.6 | 38 | 7.3 | 2 | 0.4 | 1 | 0.3 |
| | | **Adobe Sensei*** | 23 | 3.5 | 0 | 0.0 | 0 | 0.0 | 0 | 0.0 |
| | | **Canva AI*** | 1 | 0.2 | 2 | 0.4 | 1 | 0.2 | 39 | 11.3 |
| | | **Freepik** | 1 | 0.2 | 2 | 0.4 | 2 | 0.4 | 0 | 0.0 |
| | | **Adobe Photoshop*** | 0 | 0.0 | 2 | 0.4 | 1 | 0.2 | 70 | 20.3 |
| | | **FigJam AI*** | 0 | 0.0 | 74 | 14.2 | 0 | 0.0 | 0 | 0.0 |
| | | **Mural AI** | 0 | 0.0 | 1 | 0.2 | 0 | 0.0 | 0 | 0.0 |
| | | **Whimsical AI*** | 0 | 0.0 | 12 | 2.3 | 1 | 0.2 | 1 | 0.3 |
| **Video** | AI-Native Video Generation | **Hailuo AI** | 1 | 0.2 | 0 | 0.0 | 0 | 0.0 | 0 | 0.0 |
| | | **Higgsfield** | 1 | 0.2 | 0 | 0.0 | 1 | 0.2 | 0 | 0.0 |
| | | **Sora** | 1 | 0.2 | 0 | 0.0 | 1 | 0.2 | 0 | 0.0 |
| | | **Veo 3** | 1 | 0.2 | 0 | 0.0 | 0 | 0.0 | 0 | 0.0 |
| | | **Kling AI** | 0 | 0.0 | 0 | 0.0 | 1 | 0.2 | 0 | 0.0 |
| | | **Runway ML*** | 0 | 0.0 | 0 | 0.0 | 25 | 5.4 | 15 | 4.4 |
| | Video Editing Software, Embedded AI | **Captions** | 0 | 0.0 | 0 | 0.0 | 0 | 0.0 | 1 | 0.3 |
| | | **Descript*** | 0 | 0.0 | 0 | 0.0 | 0 | 0.0 | 12 | 3.5 |
| | | **Screen Studio** | 0 | 0.0 | 0 | 0.0 | 0 | 0.0 | 1 | 0.3 |
| **3D Geometry** | AI-Native 3D Generation | **Tripo AI** | 1 | 0.2 | 0 | 0.0 | 0 | 0.0 | 0 | 0.0 |
| | CAD Platform, Embedded AI | **Autodesk Generative Design*** | 11 | 1.7 | 0 | 0.0 | 0 | 0.0 | 0 | 0.0 |
| | | **Onshape** | 1 | 0.2 | 0 | 0.0 | 0 | 0.0 | 0 | 0.0 |
| | | **Siemens NX** | 1 | 0.2 | 0 | 0.0 | 1 | 0.2 | 0 | 0.0 |
| | | **Altair Inspire** | 0 | 0.0 | 0 | 0.0 | 1 | 0.2 | 1 | 0.3 |
| | | **Autodesk Fusion 360 AI*** | 0 | 0.0 | 5 | 1.0 | 9 | 1.9 | 0 | 0.0 |
| **Agentic** | Multi-step Autonomous Execution | **Marvin** | 0 | 0.0 | 0 | 0.0 | 0 | 0.0 | 1 | 0.3 |
| | | **Manus AI** | 0 | 0.0 | 1 | 0.2 | 0 | 0.0 | 0 | 0.0 |

***Note: n = number of times each tool was selected within that phase. % = n ÷ that phase's total tool selections (Discover = 664, Define = 522, Develop = 462, Deliver = 344; Table 5), not the number of participants, since one participant could select multiple tools. AI tools offered directly as preset checkbox options within phases (n = 18, 24.7%) are marked * in the table.***

## Appendix Table 4: Tool Identification Coverage for AI Adopters, by Phase

| Phase | Adopters | Linked to valid AI-Tool | | No valid AI-Tool | |
|---|---|---|---|---|---|
| | n | n | % | n | % |
| **Discover** | 318 | 314 | 98.7 | 4 | 1.3 |
| **Define** | 290 | 288 | 99.3 | 2 | 0.7 |
| **Develop** | 231 | 215 | 93.1 | 16 | 6.9 |
| **Deliver** | 182 | 178 | 97.8 | 4 | 2.2 |